\documentclass[sn-mathphys]{sn-jnl}

\jyear{2022}%

\theoremstyle{thmstyleone}%
\theoremstyle{thmstyletwo}%

\theoremstyle{thmstylethree}%

\begin{document}

\title[PINN for Dynamical PDEs]{PINN for Dynamical Partial Differential Equations is Not Training Deeper Networks Rather Learning Advection and Time Variance}


\author{\fnm{Siddharth}\sur{ Rout}}\email{sidrout@mail.ubc.ca}

\affil{\orgdiv{Department of Mechanical Engineering}}
\affil{ \orgname{The University of British Columbia}, \city{Vancouver}, \state{BC}, \country{Canada}}


\abstract{The concepts and techniques of physics-informed neural networks (PINNs) is studied and limitations are identified to make it efficient to approximate dynamical equations. Potential working research domains are explored for increasing the robustness of this technique for the solvability of partial differential equations. It is identified that PINNs potentially fails to stronger advection and longer time duration. Also, optimization function and constraint posing needs to be smarter. Even a shallow network is good for a lot of problems while powerful deeper network fails. Reservoir computing based recurrent neural network architecture is recommended to solve dynamical problems. }




\maketitle

\section{Introduction}\label{sec1}

The solution of differential equations have been an important topic for the almost every field of the world, say it be finance, mechanics, meteorology etc. Starting from the days of Newton and Leibniz solving differential equations have been core to developments in this world. Not all differential equations are solvable by hand and initiate limitations, especially when multiple independent variables come into the equation or build system of equations. However, solving these equations are need of the hour and hence we shifted our focus from exact solutions to approximate solutions as function operations and transformations are limiting in concepts. There arose the generation for solving equations by using the basic definitions or first principles of limits, discretization and numerical analysis\cite{1}. Popular such techniques are finite element methods, finite volume methods, finite difference approximations etc. These methods are generalizable and that is the benefit. Using these techniques we can solve almost any equation for any geometry. But with increased complexity, there are a couple of problems accompanied like how good the approximation is and computation expense. Discretization gives us a long list of simplified approximate equations to solve. Though we know how to solve however solving them will require us to use computers to do those hectic mathematical calculations. For a stable and accurate solution, a lot of time and energy are consumed in the process\cite{1}. Though we can calculate a lot of things in this world but we have lack of time, resources, money and problems could be endless. A type of differential equation called dynamical systems is tough to solve after a certain range in time, there are reasons to it. Dynamical systems are tough to solve as the solution my bifurcate and that shall make the system chaotic. In a chaotic system, a minute change in the initial condition or equation coefficients will have drastically different outcomes. This is sometimes referred by 'The Butterfly Effect'. The aim of this project is to develop a function approximation method that can potentially replace computationally expensive solvers for dynamical systems. The one dimensional Kuramoto-Sivashinsky equation is solved for trials and research\cite{2,3}.    

Function approximations or analytical solutions are better known for their light-weight\cite{4}. These techniques can get rid of the three primary types of errors that are evident in full order discretized approximation, namely instability, inaccuracy and shift\cite{5}. A system of dynamical systems is mainly sensitive inaccuracy due to insanely evident sensitivity to initial conditions. A benefit of function approximation is the ability of correction and reproduction. Also, an added benefit is extrapolability. Among the function approximators, neural networks have been excellent candidate as universal approximators\cite{6,7}. In the past decade, deep neural networks which are basically multiple layers of neural networks have been used for various complex regression problems due to its ability to capture high dimensional strong non-linearity. These models are fitted to the data directly as input to output mapping\cite{7}. Neural networks can also be trained to differential equations by optimizing the residuals after fitting to random points in the input domain. Such models are called physics-informed neural networks or popularly called PINNs\cite{4,5,8,9}.

Solving partial differential equations using PINNs are universally accepted by the scientific community. There are a plenty of advantages of these methods over conventional methods. The major once are ability to solve wide category of problems that were tough to solve otherwise. Moreover, these methods do not require meshing and discretization which is sometimes a tough task. Another advantage unlike other analytical models does not require data from full order solutions to set the parameters. However, being newly developed these techniques are not robust enough for solving complex equations like hyperbolic equations, strongly non-linear equations, strongly advective equations\cite{5}, chaotic dynamical systems, coupled system of equations, shock wave equations etc.

\section{Dynamical Partial Differential Equations}\label{sec2}
 Activities in the world is mostly the four dimensions, three in space and one in time. Each new dimension adds a layer of complexity. Dynamical systems generally mean functions that describe the dependence of state of a system with time. Henri Poincare was the first one to identify the special behaviour of dynamical systems. The theory of these dynamical systems is highly relevant in studying behaviour of complex dynamics, usually in the form of differential equations, which makes it continuous dynamical systems. The major points of focus in this domain are the attractors, chaos, fractals and bifurcations that explains the long term behaviour of states qualitatively. This helps understanding evolution of dynamical events like turbulence, storm, mixing fluids, environment change, economic changes, planetary motions and many more. 

 The necessary applications of dynamical systems theory are to find structural stability, Lyapunov time, bifurcation points, position tracking and quantitative approximations which one way or the other determines the predictability of the state at a particular time. Predictability of dynamical systems is a tough job. Before the advent of computing machines prediction required sophisticated mathematical techniques that were specific to specific classes of dynamical systems. These are sometimes among the toughest differential equations to solve. Also considering other factors mentioned above, accurate prediction is a great deal for these kinds of systems.

 \section{Case Selection}\label{sec3}
 The cases below point out two major difficulties in solving differential equations clearly. The concepts are explained with reference to the terms and framework of the equation mentioned. The two equations shall be good examples to analyse the theory of PINNs.
 
 \begin{figure}
    \centering
    \includegraphics[width=6cm]{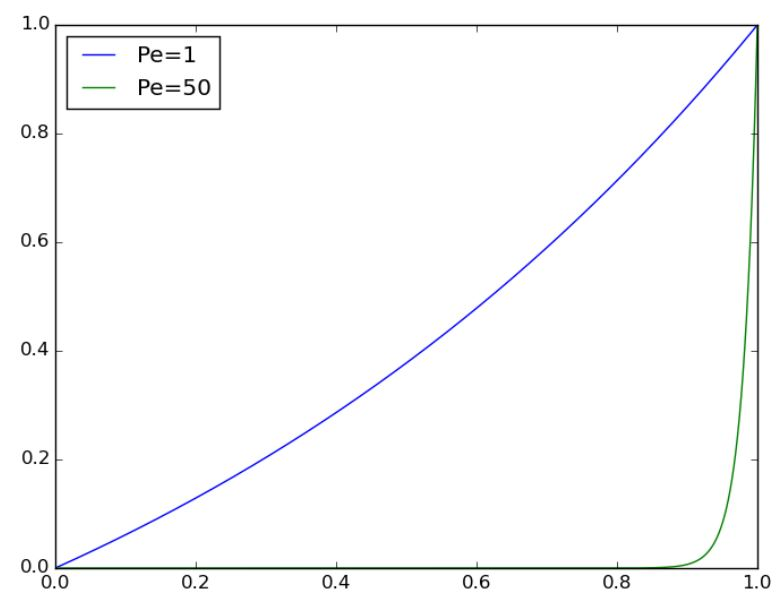}
    \caption{Solution of steady state advection-diffusion}
    \label{fig:cd}

    \raggedright
    \includegraphics[width=6.4cm]{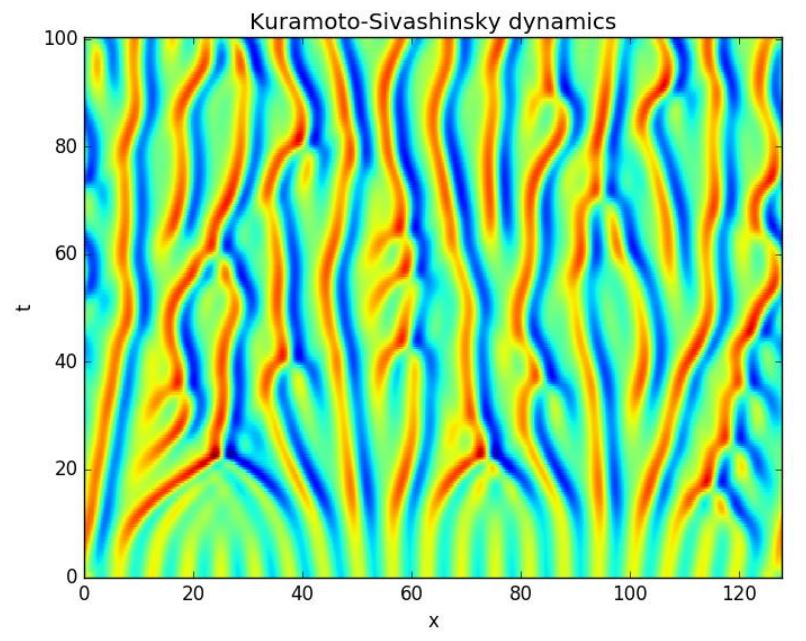}
    \caption{Solution of One dimensional Kuramoto-Sivashinsky equation}
    \label{fig:ks}
\end{figure}

\subsection{1D Steady Advection-Diffusion Equation}\label{subsec2}
The differential equation below is the governing equation for steady one dimensional flow of combined advection and diffusion phenomena. 
$$
\alpha u_{x} = u_{x x}
$$
If we notice, $\alpha$ is the weight for the advection term in the equation. That means larger is $\alpha$, more dominant is the advection effect, which introduces directional characters and hence discrete approximation becomes tougher. Figure 1 shows the difference in the solution with advection dominance. Higher is the Peclet number, more dominant is the advection. The figure compares the solution for Pe 1 and 50. Hence the numerical integration sees rapidly growing error that makes the solution unstable and inaccurate\cite{10}. Hence, a major class of higher order methods are developed to tackle this particular issue.

\subsection{1D Kuramoto–Sivashinsky Equation}\label{subsec2}
The equation below is the one dimensional Kuramoto-Sivashinsky equation
$$
u_{t}+\alpha u u_{x}+\beta u_{x x}+\gamma u_{x x x x}=0
$$
The linear form of it is as below:
$$
u_{t}+\alpha u_{x}+\beta u_{x x}+\gamma u_{x x x x}=0
$$
This equation has the advection, diffusion and dissipation effect. It is one of the equations where the solution is extremely sensitive to the initial condition\cite{2,3}. The higher order terms in the expansion of the difference equation are very much relevant and hence sensitive for error propagation in time. Figure 2 shown the solution of a case of one dimensional KS equation on Julia code developed by Mahatab Lak et. al. from the University of New Hampshire.

\section{Neural Networks as Universal Function Approximator}\label{sec4}
George Cybenko was the one to prove arbitrary width case using neural networks with sigmoid activation in 1989\cite{6}. Later in the same year, Hornik et. al. proved multi-layer feed-forward networks are universal approximators\cite{7}. Multi-layer artificial neural networks are composites of weighted sum of inputs passed through non-linear(activation) functions like tanh(), sigmoid(), etc. This enables an extremely potent highly non-linear function with large number of trainable parameters(weights and biases). This makes it universal approximation.

\section{Physics-informed Neural Networks}\label{sec5} 
Informing the physics to a neural network is a concept brought up by Lagaris et. al\cite{4}. in the late 1990s by using neural network as a trial function to solve differential equations by reduction of the residuals using AutoGrad (an automatic differentiation technique) at various points in the domain. The boundary constraints are forced into the neural network function by modifying it mannually. In 2017, Raissi et al. proceed by using more accurate automatic differentiation and deeper networks to approximate tougher problems\cite{8,9}. The novelty in their work comes from the way they pose the loss function to reduce the residual. They did not manually force the constraints by modifying the trial function rather let the trial function fit to the boundary and initial constraints by adding the mean square error from the data points satisfying the conditions as summed constraints to mean squared residuals. This makes the technique very much generalizable. Almost all the differential equations could be posed to be solved using this technique, which they named PINNs. 

\subsection{Advancements in PINNs}\label{subsec2}
The unknown solution $\boldsymbol{u}(t, \boldsymbol{x})$ is represented by a deep neural network $\boldsymbol{u}_{\boldsymbol{\theta}}(t, \boldsymbol{x})$, where $\boldsymbol{\theta}$ denotes all tunable parameters of the network (e.g., weights and biases). The physics-informed model can be trained by minimizing the following loss function.

$$
\mathcal{L}(\boldsymbol{\theta})=\lambda_{i c} \mathcal{L}_{i c}(\boldsymbol{\theta})+\lambda_{b c} \mathcal{L}_{b c}(\boldsymbol{\theta})+\lambda_{r} \mathcal{L}_{r}(\boldsymbol{\theta}),
$$

where

Here $\left\{\boldsymbol{x}_{i c}^{i}\right\}_{i=1}^{N_{i c}},\left\{t_{b c}^{i}, \boldsymbol{x}_{b c}^{i}\right\}_{i=1}^{N_{b c}}$ and $\left\{t_{r}^{i}, \boldsymbol{x}_{r}^{i}\right\}_{i=1}^{N_{r}}$ can be the vertices of a fixed mesh or points that are randomly sampled at each iteration of a gradient descent algorithm. The hyper-parameters $\left\{\lambda_{i c}, \lambda_{b c}, \lambda_{r}\right\}$ allow the flexibility of assigning a different learning rate to each individual loss term in order to balance their interplay during model training\cite{12,13}. These weights may be user-specified or tuned automatically during training.

\subsection{Advantages of PINNs over Other Neural Networks}\label{subsec2}
The major advantage of this technique is that it does not require physical data to train the analytical model. Moreover, the technique is generalizable in the sense that with the exactly same concept various equations could be solved\cite{4,5,8,9,12,13}. Previous models required alteration of the learning function depending on number of equations coupled, boundary conditions etc. in order to force the constraints. Being a strong approximating function the three major kinds of error, namely instability, inaccuracy and shifting errors, could be taken care of simultaneously. These problems are taken care individually in finite numerical techniques\cite{5}. This kind of technique is an excellent candidate for robust higher order methods. Neural network is not just an approximator but rather a smart approximator. Hence, depending on the local physical property it can act differently with switching kind of behavior. In particular for the case of dynamical system, with time progression the integrated error increases too rapidly. PINNs being an optimization technique for regression, training to measured physical data points as additional loss terms or regularization can be used for correcting the approximating function.

\section{Experiments with the selected cases}\label{sec6} 
The qualitative property in dynamical problems is strong translational variance. Hence, the two major causes for PINNs performing poorly are advection dominance and time variance which are demonstrated below.
\subsection{PINNs for 1-D Steady Advection-Diffusion}\label{subsec6}
Peclet number is a good non-dimensional parameter to scale advective dominance over diffusion characteristic of an equation. It is the ratio of advective transport rate to diffusive transport rate. In our problem we can quantify that by the ratio of coefficient of advective term times the length of domain space to coefficient of diffusive term, so that is $\alpha$ in our particular case. PINN can solve this problem but there is a limit set by advection characters in the differential equations. No matter how deeper and how sophisticated we make the neural network it is not possible to solve for problems with Peclet number more than something close to 8. Figure 3 shows the results noted. For lower Peclet number problems, it is noticed that it is not necessary to have deeper and wider layers. A good thing is in conventional numerical techniques fails for problems set with this value more than 2. Hence, these schemes can be used for shape functions that let larger grids with similar accuracy. The work and figures in this section has been sourced from by 2019 thesis titled "Numerical Approximation in CFD Problems Using Physics Informed Machine Learning"\cite{5}.\\
\begin{figure}
    \centering
    \includegraphics[width=6cm]{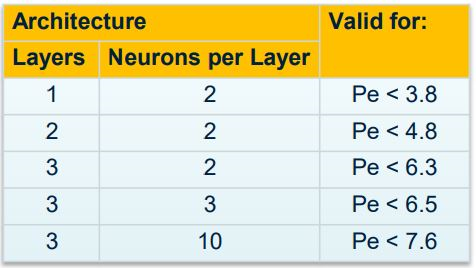}
    \caption{Optmimized architecture for solving one dimensional steady advection diffusion.}
    \label{fig:cdPinn}
\end{figure}
\begin{figure}
    \centering
    \includegraphics[width=6cm]{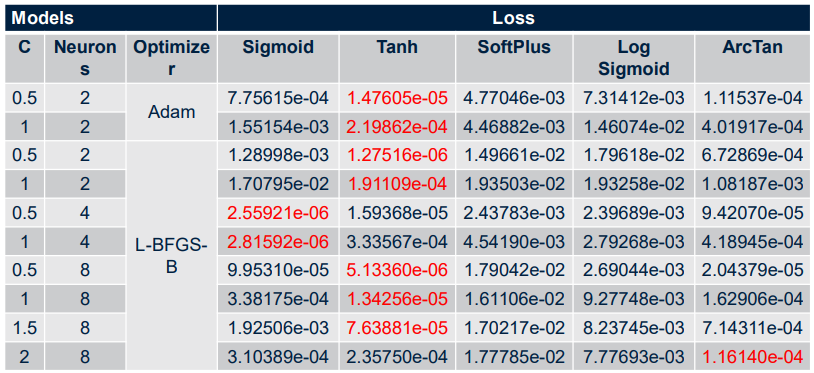}
    \caption{Minimized losses with various activation functions while solving one dimensional steady advection diffusion where C is coefficient of advective term.}
    \label{fig:cdPinnAf}
\end{figure}
 \begin{figure}
    \centering
    \includegraphics[width=8cm]{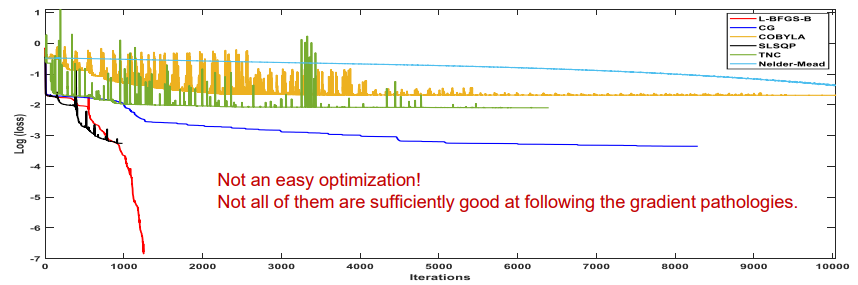}
    \caption{Loss trend while solving using various optimizer for one dimensional steady advection diffusion}
    \label{fig:cdPinnOpt}
\end{figure}
Parametric analysis is done to understand more about the performance and effectiveness. The impact of change in the non-linearity function as well as loss optimization algorithm is studied. Figure number 4 records the loss values in a tabular form. Among various non-linearity functions tanh() and tan() performs consistently as well as better than sigmoid(). However, tanh() wins this game clearly. Figure number 5 shows the trend of loss value with iteration for various optimizers. The neural network is trained using various optimizers however L-BFGS-B and SL-SQP perfoms better than other specialised optimization techniques. BFGS performs remarkably better than others. The prime reason could be the fact that these optimizers are second order and perform better in the case of optimizing multiobjective functions than other techniques. However, it can be noted that first order techniques like Adam performs decently even if they don't fail. An important thing that can be observed that the collocated PDE residuals and the fitting losses for constraints are clearly not in similar scale which is not typical for regular data-driven neural network learning. Also the gradient pathology is not smooth and hence tough for other optimizers. Guided optimization like hill climbing helps in few cases.

\subsection{PINNs for 1-D Kuramoto-Sivashinsky}\label{subsec6}
There are recent publications demonstrating how PINNs can solve one dimensional Kuramoto-Sivashinsky equation which is among the standard dynamical equation that can turn chaotic. However, there is a small intersecting set of coefficients for advection, diffusion and dissipation terms for which this works. As demonstrated in the previous case the dominance of advection toughens the optimization of the PINN. Here, not only it is dynamical but also non-linear. It has various orders of spatial derivatives making it difficult especially when the system becomes chaotic.\\

CausalPINN by Wang et. al. is considered the state of art PINN\cite{13}. They have rightly identified the problem of multiobjctive optimization due to difference in scales of residuals and constraints stated by Rout et. al.\cite{5} and devised weights for each loss terms by normalizing with the each cumulative loss terms. They are first people to be able to solve 1D KS Equation. We can validate their model using their open source code provided. Figure 6 shows how CausalPINN perfoms with time for the case provided in the figure. It can be noticed that PINN can now solve complex dynamical problems. The initial sine smoothly curls as expected while the constraints are obeyed. Like initial curve is a neat sine function and the boundaries are continuously at base zero(0). However, the typical equation where all the coefficients are considered 1 is taken for regular study of the equation. The state of art PINN fails to optimize even after an effort equivalent to one-day's run-time. The net loss is recorded to be 33.263 where the constraint loss was 0.0016. This suggests the difficulty in fitting to the PDE where as PINN could manage to obey the constraints. The residual loss is noted to be 1808.506, where its weight in the loss function disappear from the scale. This clearly shows the issue of multi-objective optimization.

\begin{figure}
    \centering
    \includegraphics[width=7.5cm]{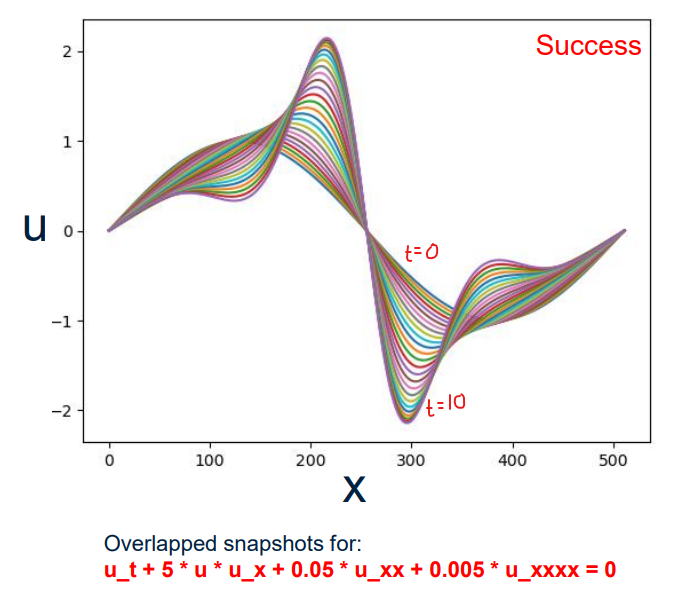}
    \caption{Snapshots of velocity(U) along the x-axis overlapped through time for one dimensional Kuramoto-Sivashinsky equation.}
    \label{fig:KSPinnPlt}
\end{figure}

\section{Observations and Conclusion}\label{sec7}
Based on the experiments and analysis a few points could be commented. Simple addition or weighed by mean addition of squared loss terms of residuals at collocation points and fitting points directly for constraints have different scale and orders of magnitudes. It is one of the major issue as it can lead to non-pareto optimal solution. It is also noticed that sometime the loss function gets stuck at local optima and gradient pathologies are tough and uneven in the parameter hyperspace\cite{5,12}. Hence, stochastic first order optimizers work otherwise higher order optimizers suitable for constrained optimizations like SQP and BFGS works while other fails\cite{5}. A better representation of loss function like appropriate or adaptive weighted losses can help. Otherwise constraints could be forcefully enforced by modified architecture or trial function, like explained by Isaac Lagaris et. al\cite{4}. Specifically, in the context of problems in dynamical systems recurrent neural networks(RNN) could prove to be better candidate over simple deep networks\cite{14}. A concrete reasoning has been provided by Eldad Haber et. al. where RNNs can be proved to be in the form of differential equations and hence fit into the theory to learn the dynamical differential equations better\cite{15,16}. Especially, for chaotic systems reservoir computing have been proved to be performing better\cite{11}.
Reservoir computers are a class of RNNs where the intermediate nodes are randomly arranged and connected\cite{11, 14}. They have random recurrent connections. The intermediate nodes are jumbled and entangled however they are connected out to the output layer linearly. The entangled architecture makes it tough to backpropagate and hence only the final layer of weights are trainable for convenience. The trainable output layers makes the effective non-linear network linear with respect to trainable parameters hence conserving strong non-linearity while making it easy to train. Apparently, RNN can also be introduced with PINNs kind of loss definition to solve chaotic problems for turbulence and extreme event prediction\cite{17}.\\
Ultimately, we can justify the errors and specify the right path to solve a dynamical system of partial differential equations by identifying the two prime cause of poor performance. The two causes are advection dominance and time variance, which have been identified from the case studies. We can conclude that there is not always a requirement for deep and bulky layers in the architecture. The criticality lies in the way it is posed for optimization and the optimizability. Gradient pathology must be taken care of. Deeper layers give the potential to capture extremely strong non-linearity in high dimensional and strongly coupled system of equations. Also, specifically for time variance characteristic we should use recurrent neural networks, especially reservoir networks which are in fact light weight but performs better. "Physics-Informed Recurrent Neural Networks (PIRNN) is the right path for solving dynamical and chaotic problems".         


\bibliography{sn-bibliography}

%

\end{document}